\documentclass[12pt]{article}

\usepackage{epsfig}

\begin{document}

\title{\bf{\Large\boldmath \unboldmath 
Influence of dipole interaction on lattice dynamics of 
crystalline ice}}
%
   \author{\normalsize
W.A.\ ADEAGBO\footnote{
      Corresponding author. Tel.\ +49 203 379 1606;
      fax: +49 203 379 3665. 
     \newline E-mail: adeagbo@thp.uni-duisburg.de}
\ and P.\ ENTEL 
\\*[0.2cm]
{\small \it Institute of Physics,} \\
    {\small \it University of Duisburg-Essen, Duisburg Campus, 47048
     Duisburg, Germany}\\
%
\\*[0.2cm]
}
\date{\small \it (Received  \today )}
\maketitle

\begin{abstract}
%
\noindent
The Born effective charges of component atoms and phonon spectra 
of a tetrahedrally coordinated crystalline ice are calculated 
from the first principles method based on density 
functional theory within the generalized  gradient approximation with the 
projected augmented wave method. Phonon dispersion relations in a 
$3\times1\times1$ supercell were evaluated from Hellmann-Feynman forces with 
the direct method. This calculation is an additional work to the direct method 
in calculating the phonon spectra which does not take into account the 
polarization charges arising from dipole interaction of molecules of water 
in ice. The calculated Born effective polarization charges  
from linear response theory are supplied as the correction terms to the 
dynamical matrix in order to further investigate the LO-TO splitting of the 
polar modes of ice crystal at ${\bf k}=0$  which has long been speculated 
for this system especially in the region between 28 and 37 meV both in the 
theoretical and experimental studies. Our results clearly show the evidence 
of splitting of longitudinal and transverse optic modes at the 
${\bf k}=0$-point in agreement with some experimental findings.
\end{abstract}

\section*{1. INTRODUCTION}
The vibrational studies of solid ice have gained attraction for scientific 
investigation over the decades. Experimentally, the vibrational spectra of 
the different phases of ice have been investigated using infra-red (IR) and 
Raman techniques, but due to the proton disordering in most such structures, 
the normal selection rules governing the interaction of radiation with these 
lattices are broken and hence any analysis of the spectra is difficult. 
On the other hand, the IR and Raman spectra are very sensitive to the 
intramolecular modes involving O-H stretching and bending, and less 
sensitive to the intermolecular modes involving the vibrations 
where the whole water molecules are moving against each other. 
Therefore, in normal circumstances, only limited information 
(the acoustic frequencies in 
particular) can be obtained in the translational region. Inelastic incoherent 
neutron scattering (IINS) is only more reliable because its spectrum is 
directly proportional to the phonon density of states weighed by mean square 
amplitude associated with each mode and also the selection rule is not 
involved as all modes are measured 
simultaneously~\cite{Li1994,Li1996}. Despite the detail information 
 that can be obtained from the present days methods of coherent and 
incoherent scattering on the vibrational motions of the atoms or molecules 
of most systems based on the  peak positions, intensities and their width, 
there  are still  many problems about the ice  systems.  These problems as 
mentioned, are usually associated with the the proton disordering. 
The phonon spectra measurements of Renker~\cite{Renker1973} on Ice-Ih 
(D$_2$O) made by time of flight using a chopper spectrometer were extensive 
but not complete in that information above 20 meV in the (001) direction and 
above 30 meV in other directions is missing, presumably because of a lack of 
scattering intensity and of the nature of the disordering of the protons in 
the ice structure.  \\
\indent
First-principles calculation of crystal structure and crystal 
properties is becoming standard technique, and the progress in the methods, 
algorithms, and computer capabilities allows to study larger systems of 
solids crystals in which crystalline ice cannot be left out.
The method has recently gained ground not only because of its reliability 
in the study of static and dynamics properties of 
ice~\cite{ Cote2003,Morrison1999} but also some important
features such as modeling ordered periodic ice structure~\cite{Lee1993},
and also to probe the nature of hydrogen bond in different 
geometries~\cite{ Xantheas1993}. Its theoretical counterpart such as the 
classical modeled potential through empirical 
method~\cite{Massimo1986,Glen1984} had some success in describing 
some important dynamical features, but to date, there is none 
capable of describing the ice dynamics and related properties across its
whole spectral range and describing certain key spectral features.\\
\indent
There are two techniques currently in use in the first-principles method 
in the study of lattice dynamics of crystals: The linear response method and 
the direct method. In the linear response method the dynamical matrix is 
obtained from the modification of the electron density, via the inverse 
dielectric matrix. The dielectric matrix is calculated from the 
eigenfunctions and energy levels of the unperturbed 
system~\cite{Resta1985}. It can be determined at any wave vector
 in the Brillouin zone with the computational effort required comparable to 
that of a ground state optimization. Only linear effects, such as 
harmonic phonons, are accessible to this technique. 
On the other hand, the direct-method is based on the solution of the 
Kohn-Sham equation and it allows one to study both linear and 
non-linear effects. The calculations deal with a supercell, which allows 
explicit account to be taken of any perturbation. This method is rather 
straightforward computationally and there are a few standard software 
packages. Within the direct method the phonon 
frequencies are calculated from Hellmann-Feyman forces generated by the 
small atomic displacements, one at a time. Hence using the information of 
the crystal symmetry space group the force constants are derived, and the 
dynamical matrix is built and diagonalized, and its eigenvalues arranged into 
phonon dispersion relations. In this way, phonon frequencies at selected high 
frequencies at high-symmetry points of the Brillouin zone can be 
calculated~\cite{Parlinskin2002}.    
However, when the interaction range ceases to be within the supercell, 
phonons at all wave vectors are determined exactly.
The above statement has to be modified for polar crystals for which 
the macroscopic electric field splits off the infrared-active optic modes.
The long-range part of the Coulomb interaction corresponds to the 
macroscopic electric field arising from ionic displacements.\\ 
\indent
Ice is a tetrahedrally covalently bonded polar system whose dipole-dipole 
interactions give rise to  the electric field when
 they are  disturbed. The origin of the splitting is
 therefore the  electrostatic field created by long wavelength modes of
 vibrations in such crystals. Usually a microscopic electric field influences
 only the LO modes  while TO modes remain unaltered. The field therefore
 breaks the Born-von  K\'arm\'an conditions, as a consequence with a direct
 method only finite wave  vector ${\bf k}\ne 0$ calculations are
 possible. The LO/TO splitting has therefore been found by calculating the 
effective Born charge tensor and electronic dielectric constant introduced 
into the dynamical matrix in the form of a non-analytical 
term~\cite{Zhong1994} or by calculating LO modes from elongated 
supercells~\cite{Parlinskin1999} as will be discussed below.\\ 
\indent
In our previous work, we have applied with success the direct method to the 
calculation of phonon dispersion of ice and its corresponding vibrational 
density of states~\cite{Adeagbo2005}. The method reproduces the 
important features in  the translational mode, librational mode, 
bending as well as the stretching region in comparison to the experimental 
results. The only ingredient that was missing in the previous work has been 
explained above, i.e., a macroscopic electric field arising from dipolar 
interaction which is not taken into account. Despite the huge computational 
demand of this problem, we  did the additional calculation of Born-effective 
charges  which is supplied as ingredients to the direct method to identify the 
long-range part of interatomic force constants and makes the interpolation 
of phonon frequencies tractable. Our overall aim is to help resolve the 
discrepancies in the reported phonon frequencies especially the puzzle 
behind the LO/TO splitting of some optical modes at ${\bf k}=0$ and provide 
first principles Born-effective charges and dielectric tensors for direct 
method phonon calculations.\\ 
\section*{2. METHOD OF CALCULATION} 
\noindent 
The calculation of phonon dispersion relations were performed with the direct 
 method. The direct method uses the Hellmann-Feymann (HF) forces calculated for
 the optimized supercell with one atom displaced from equilibrium position,
 derived from the force constants using the symmetry elements of the space
 group of the crystal, and calculates phonon frequencies by diagonalizing the
 dynamical matrix.\\
In this work, the ice crystal structure optimization and calculation of HF 
forces have been performed with the density functional theory  using the PAW 
method within  the generalized gradient approximation (GGA), 
as implemented in the Vienna {\it Ab Initio} Simulation Package 
(VASP)~\cite{Kresse96} software.
A unit cell of ice crystal was prepared in a cubic box according to
Fig.~\ref{cellice} with 8 molecules of water. All the atomic degrees
of freedom  were relaxed with high precision. The optimum Monkhorst 
Pack of $4\times4\times 4$ $k$-point was used in addition to the GGA of
Perdew-Wang to describe the exchange-correlation  and the hydrogen 
bonding of water.  We used a energy cut-off
of 500 eV because the 2p valence electrons in oxygen  require a large
plane wave basis  set to span the high energy states described by the
wavefunction  close to the oxygen nucleus, and also the
hydrogen atoms  require a larger number of planes waves in order to 
describe localization of their charges in real space.
The relaxed geometry for the unit cell from the initial configurations 
containing 8 molecules is shown in Fig.~\ref{cellice}. 
The starting geometry of the molecules in the cubic simulation box shown 
is such that no hydrogen bonds were present but the
positions of oxygen atoms follow the tetrahedral geometry. 
After the relaxation, all the protons perfectly point to the right 
direction of oxygen atoms and make the required hydrogen bonds necessary  
as indicated  by the dotted lines in Fig.~\ref{cellice} to 
preserve the tetrahedral orientation of the ice structure. This final structure is in 
accordance to Bernal Fowler's rules~\cite{Bernal1933} which are 
based on the statistical model of ice. \\
The relaxed geometry is tetragonal with calculated lattice parameters
$a$ = 6.1568 {\AA}, $b$ = 6.1565 {\AA}, $c$ = 6.0816{\AA} i.e. with 
${c}/{a}$ ratio $\approx$ 0.988. The experimental lattice constant reported 
by Blackman {\it et. al.}~\cite{Blackman1958} is  6.35012652 {\AA} 
for the cubic  geometry. The calculations of force constants 
was carried out by considering a $3\times1\times1$ supercell 
containing 24 molecules of water which is obtained by matching 
3 tetragonal unit cells. At the first step of the calculation, the
PHONON  software is used to define the appropriate crystal supercell 
for use of the direct method. The phonon frequencies $\omega({\bf k},j)$ 
are calculated as square roots of eigenvalues of the supercell dynamical 
matrix:
\begin{eqnarray}
\mathrm{{\bf D^{SC}}}({\bf k})e({\bf k},j) & 
= & \omega^2({\bf k},j)e({\bf k},j),
\label{dyna1}
\end{eqnarray}
where the $e({\bf k},j)$ are the polarization vectors. The supercell 
dynamical matrix is defined as 
\begin{eqnarray}
\mathrm{\bf D^{SC}}({\bf k},\mu\nu) & = & \frac{1}{\sqrt{M_\mu M_\nu}} 
\sum_{m\in SC} \mathrm{\bf \Phi^\mathrm{\bf SC}}(0,\mu;m,\nu) \\ \nonumber  
& \times & exp(-2\pi{\bf ik}.[{\bf R}(0,\mu)-{\bf R}(m,\nu)]),
\label{dyna2}
\end{eqnarray}
where the  summation over $m$ runs over all atoms of the supercell; $M_\mu$, 
$M_\nu$ and ${\bf R}(0,\mu)$, ${\bf R}(m,\nu)$ are atomic masses and 
equilibrium vectors, respectively; and ${\bf k}$ is the wavevector. 
The cummulant force constants $\mathrm{\bf \Phi}_{ij}^{\bf{SC}}$ are the sums 
of terms containing the second derivatives of the ground-state energy 
with respect to the position vectors of interacting atoms $i$ and $j$. 
The HF forces in the direct  method  are derived using
\begin{eqnarray}
{\bf F}_i(n,\nu) & = & -\sum_{m,\nu,j} {\bf \Phi}_{ij}^{\bf SC}(n,\nu; m,\mu)u_j(m,\mu),
\label{ForceHF}
\end{eqnarray}
where ${\bf u}_j(m,\mu)$ is an amplitude of displacement of an atom in the
supercell specially shifted from the equilibrium position.\\
The symmetry of the supercell and the site symmetry of the
non-equivalent atoms usually considerably reduce the number of
displacements needed for reconstruction of ${\bf\Phi}_{ij}^{\bf SC}$. We are
unfortunate in our case because of the hydrogen bonding fluctuations 
which makes the crystal structure complicated. Therefore, we have to 
consider the whole 24 atoms in the supercell as independent displacements: 
In the positive and negative non-coplanar, $x$, $y$ and $z$ directions. 
As done for the 
primitive unit cell, all the internal coordinates were relaxed until the 
atomic forces were less than $10^{-4}$ eV/{\AA}.
Complete information of the values of force 
constants were obtained by displacing every atom of the primitive unit cells by 0.02 {\AA}
in both positive and negative  $x$, $y$ and $z$  directions.
Therefore, minimization of the anharmonic effects and systematic errors are achieved by calculating ${\bf\Phi}_{ij}^{\bf SC}$ with Eq.~\ref{ForceHF} 
using forces arising from both  positive and negative displacements $u_j$.
As mentioned above, we use a $3\times 1\times 1$ supercell, which implies that
3 points in the direction [100] are treated exactly according to the 
direct method. The points are [$\zeta$00], with  $\zeta$ = 1, 1/3, 2/3.
We calculate forces induced on all atoms of the supercell when a
single atom is displaced from its equilibrium position, to obtain the force
constant matrix, and hence the dynamical matrix. This is then followed by
diagonalization of the dynamical matrix which leads to a set of eigenvalues
for the phonon frequencies and the corresponding normal-mode
eigenvectors. The vibrational density of states (VDOS) is obtained by
integrating over  ${\bf k}$-dependent phonon frequencies from the 
force-constant matrix in supercells derived from the primitive
 molecule unit cells.\\
For the ionic crystals the macroscopic electric
field is taken into account by adding to Eq.~(\ref{dyna1})
the non-analytic term of the dynamical matrix at the wave vector
${\bf k}=0$~\cite{Pick1970}.
However, since one knows the phonon frequencies only at discrete wave
vectors, it is justified to extend the non-analytical term to the
${\bf k}\ne0$ region, through multiplying it by the Gaussian damping factor.
Therefore we  replace Eq.~(\ref{dyna1}) by the following expression:
\begin{eqnarray}
  {\bf D}_{\alpha,\beta}^M ({\bf k}; \mu\nu) &  = & {\bf
  D^{SC}}_{\alpha,\beta} ({\bf
    k}; \mu\nu) \nonumber  \\
  & + & \,\frac{4\pi e^2 }{ V\epsilon_\infty \sqrt{M_\mu M_\nu}} \frac
  {[{\bf k
        }\cdot{\bf Z}^*(\mu)]_\alpha [{\bf k}\cdot{\bf Z}^*
  (\nu)]_\beta}  {|{\bf
      k}|^2} \nonumber\\
  & \times & \,\mathrm{exp}  [-2\pi i {\bf g} \cdot ({\bf r}(\mu)-
  {\bf r}(\nu))
    ] \nonumber \\
    & \times & \, d({\bf q}) \mathrm{exp} \left\{ -\pi^2 \left[ \left(
  \frac
    {k_x}{\rho_x} \right)^2 +  \left( \frac
      {k_y}{\rho_y} \right)^2 + \left( \frac
        {k_z}{\rho_z} \right)^2 \right]\right\},
    \label{LOTO1}
    \end{eqnarray}
where ${\bf k}$ is the wave vector within the Brillouin zone with its
centre at the reciprocal-lattice vector ${\bf g}$, $V$ stands for the volume of
the primitive unit cell, and $M_\mu$, ${{\bf r}_\mu}$  are atomic
masses and internal positions, respectively. The ${\bf Z}^*(\mu)$ are the
tensors of the Born-effective charges. $\epsilon_\infty$ is the
electronic part of the dielectric constant and $\rho$ $(x$, $y$ and
$z$) are damping factors; then the non-analytical term vanishes close to the 
zone boundary. Consideration of the effective charges leads to the LO/TO
splitting of the optical parts of the phonon modes of of ice at the 
$\Gamma$-point as discussed in the next Section. 
This observation has long been been
speculated both from theory and experiment.\\
By definition~\cite{Rici1994}, the Born effective charge  tensor 
${\bf Z^*}_{i,\alpha\beta}$ quantifies to linear order the polarization per unit cell 
(${\bf P}$) generated by zone-center {\bf k}=0,
created along the direction $\beta$ when the atoms of sublattice $i$ are displaced 
in the direction $\alpha$ under the condition of zero electric field. It is calculated 
according to the equation:
\begin{eqnarray}
{\bf Z^*}_{i,\alpha\beta} = {\bf Z}_i + \Omega \frac{\partial \bf P_\alpha}
 {\partial \bf u_{i,\beta}}.
\label{effZ}
\end{eqnarray}
The macroscopic dielectric constant is found via the relation 
\begin{eqnarray}
\epsilon_\infty = 1 + \frac {4\pi{\bf P}}{\bf E},
\label{dielec}
\end{eqnarray}
where ${\bf E} = {\bf E}_{ext}-4\pi {\bf P}$ is the total macroscopic 
electric field.\\
\vskip 1truecm
\noindent

\section*{3. RESULTS AND DISCUSSION} 
Table \ I shows the dielectric constants  of the $\epsilon_{\infty}$ tensor
and the Born-effective charges calculated according to  Eq.~(\ref{dielec})
and~(\ref{effZ}) implemented in PWSCF$_{2.1}$ code~\cite{Baroni2001}. 
The dielectric constant tensor is symmetric with non-zero off-diagonal terms 
but with negligibly  contribution in comparison to the diagonal element.
 The diagonal term $\sim$ 1.88 is in a very good range for the high frequency 
limit (Thz) of the dielectric constant of ice~\cite{Victor1999}.
Under an applied field the individual molecules are 
polarized by the field. This involves the displacements of the electrons 
relative to the nuclei and small distortions of the molecules under the 
restoring forces. The response to a change in field is very rapid, 
so that the effects are independent of frequency  up to microwave frequencies.
 The polarization in ice in general is due to the reorientation of molecules 
or bonds, that is, the energies of some of the proton configurations, that are 
compatible with the ice rules~\cite{Bernal1933,Pauling1935}, 
are lowered relative to others, so that in thermal equilibrium there is 
net polarization of  ice. The achievement of this equilibrium state is a 
comparatively slow process that requires thermal activation and local 
violations of the ice rules.\\    
Our results for the dynamical effective charges are shown in Table I for 24  
non-equivalent atoms. The complication of the system due to the protons 
re-orientation and the hydrogen bonding make the symmetry consideration 
difficult. The charge neutrality condition~\cite{Pick1970}
 requires that the acoustic mode frequencies vanish for ${\bf k =} 0$ such that
\begin{eqnarray}
{\bf Z}^*_{i,\alpha\beta} = 0.
\label{neutralQ}
\end{eqnarray}
Therefore, this condition   is satisfied to at least order of $10^{-4}$ 
electron which is accurate enough for any reliable  calculation. 
The unequal effective charge tensor components ${Z}^*_{\mathrm H}$ and 
$Z^*_{\mathrm O}$  of each hydrogen and oxygen atoms 
$Z^*_{xx}$ $\ne$ $Z^*_{yy}$ $\ne$ $Z^*_{zz}$ are due to the broken symmetry 
arising from the lattice distortion. The values alternate among the 
component atoms of H and also for O in order to preserve the overall 
neutrality. For instance, the  $Z^*_{zz}$  of the hydrogen atoms is 
0.624($\pm$0.001) electron, while the other two elements  $Z^*_{yy}$ and 
$Z^*_{zz}$ alternate within 0.666($\pm 10^{-4} $) electron. The off-diagonal 
elements have the values which ranges within  $\pm 0.408$, $\pm 0.407$, 
$\pm 0.402$, $\pm 0.377$ and $\pm 0.382$ with deviation $\pm 10^{-4}$ electron.
Similar features are observed for the oxygen atoms with $Z^*_{zz}$ -1.070 
$\pm 10^{-4}$ electrons  while $Z^*_{xx}$ and  $Z^*_{yy}$ alternate  between  
-1.084 and -0.961 with deviation ($\pm 10^{-4}$) electrons. Some of the 
off-diagonal contributions are too small. The observed anisotropic features 
can be attributed to the complexity of the hydrogen bond during the electron 
transfer process and also due to the dipole interaction  
of the water molecules in Eq.~\ref{LOTO1}.\\
Figure~\ref{dispVDOS}  shows the dispersion relation obtained by supplying the 
calculated effective charges and the corresponding highest frequency 
dielectric constants, shown in Table I, as the correction from the 
analytical term  which was added to the dynamical matrix as explained in 
Section 2. We also compare the dispersion obtained in the absence of these 
charges to see the magnitude of splitting in the optical mode. 
The calculated VDOS for both dispersions do not appreciably change much 
because the states are not complete as it requires summation over all points 
in the first Brillouin zone. The LO modes that was formally degenerate in 
the absence of $Z^*$ at 27.1 meV in the translational region is now shifted 
to a higher value 30.2 meV as shown in Fig.~\ref{Tran}. 
Except for the isotopic effect due to the different 
masses of hydrogen and deuterium  atoms in ice (see the experimental 
dispersion on the right of  Fig.~\ref{Tran}). This observation correctly 
shows the splitting in the  optical modes due to the dipole interaction 
of the water molecules which correspondingly  induce a dipole moment in the 
optical mode. Apart from the translational region, as shown in 
Fig.~\ref{libest}, other splitting of LO modes occur at  111.0 meV in the 
librational region with a small shift to 112.5 meV.
Also, there is a small shift of 0.1 meV from 204.6 meV in the bending region. 
The large shift of about 12.0 meV is observed  from 362 meV in the stretching 
region because the strength of dipole interaction is large when there is 
symmetric stretching ($\nu_1$) of O-H bond, while a tiny effect  
observed in the antisymmetric stretching is due to the compensation arising 
from simultaneous bond lengthening and shortening of the O-H covalent bond.
\section*{4. SUMMARY}
In summary, we have performed the analysis of the lattice dynamics in 
crystalline tetrahedrally coordinated ice and found a very strong influence 
of the dipole interaction on the phonon spectra in the optical regions of ice. 
{\it Ab initio} calculations clearly show the splitting in the 
region between 27.1 and 30.2 meV of the translational region .
This observation can be correlated with the experimental observation 
using high resolution inelastic 
neutron measurements of the phonon density of states, in which two separate 
molecular optical bands at about 28 and 37 meV for ice Ih and ice Ic have been 
observed~\cite{Li1996}.  Other splitting of LO modes in the 
librational, bending and stretching region were also  predicted due to 
the dipole interactions. The calculation also shows the extent to which the 
direct method can be used to calculate the phonon spectra of the dipole 
system. To our knowledge, a strong influence of the dipole interaction 
on the lattice dynamics of ice from effective charges calculation was not yet 
reported.\\\\\\

\section*{\bf  Acknowledgments}

We acknowledge the  support by the Deutsche Forschungsgemeinschaft
(Graduate College 277 {\em ``Structure and Dynamics of 
Heterogeneous Systems''}).


\clearpage\centerline{TABLES}
\begin{list}{}{\leftmargin 2cm \labelwidth 1.5cm \labelsep 0.5cm}

\item[\bf Table 1]
Calculated dielectric constants and Born effective charges 
of cubic ice using linear response in  PWSCF$_{2.1}$ code~\cite{Baroni2001}.

\end{list}

\clearpage \centerline{FIGURE CAPTIONS}

\begin{figure}[h]
  \centering
  \resizebox{12cm}{!}{\includegraphics*{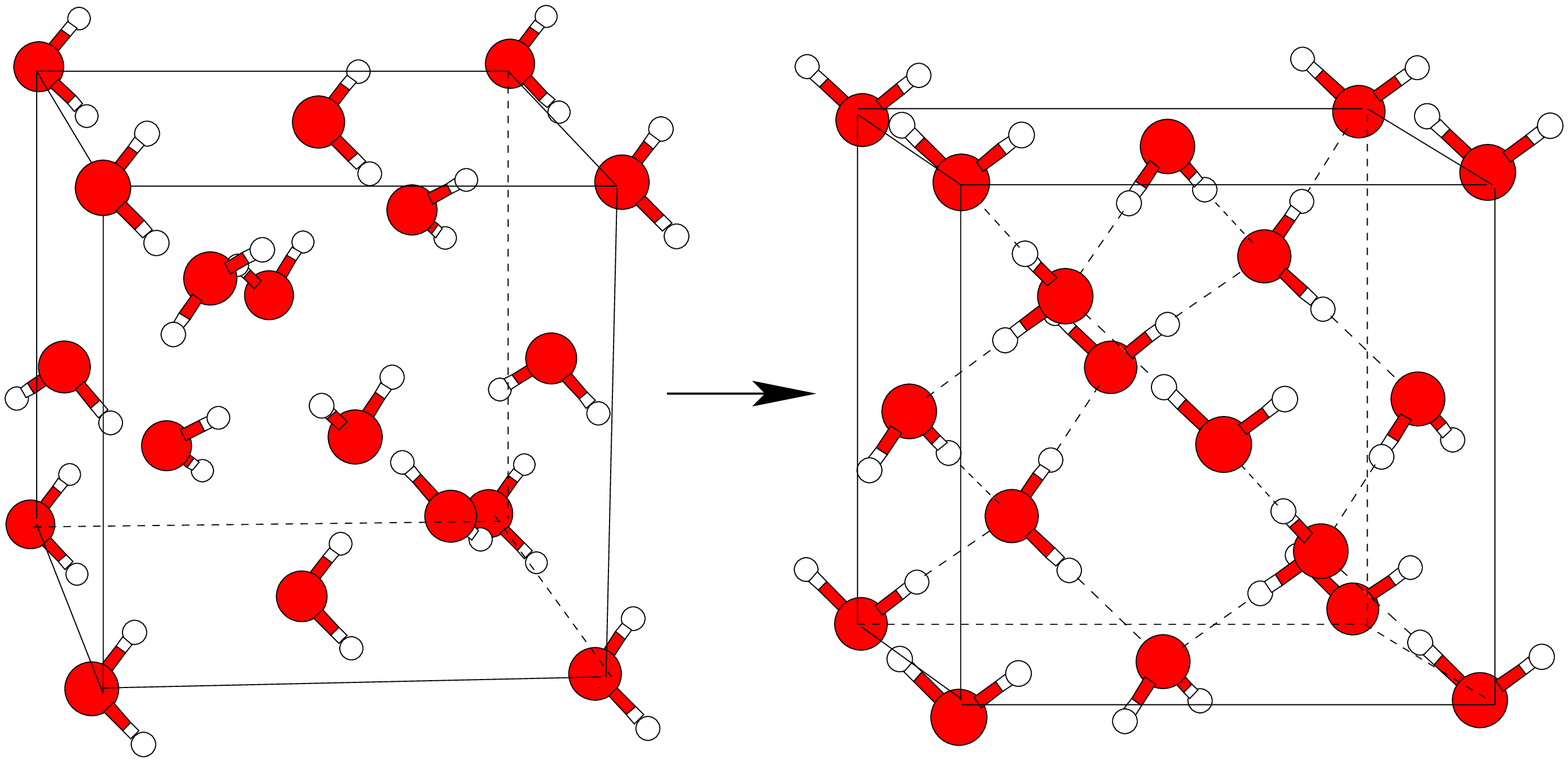}}
\caption{Adeagbo et al.}
\label{cellice}
\end{figure}

\begin{figure} 
 \centering
  \resizebox{14cm}{!}{\includegraphics*{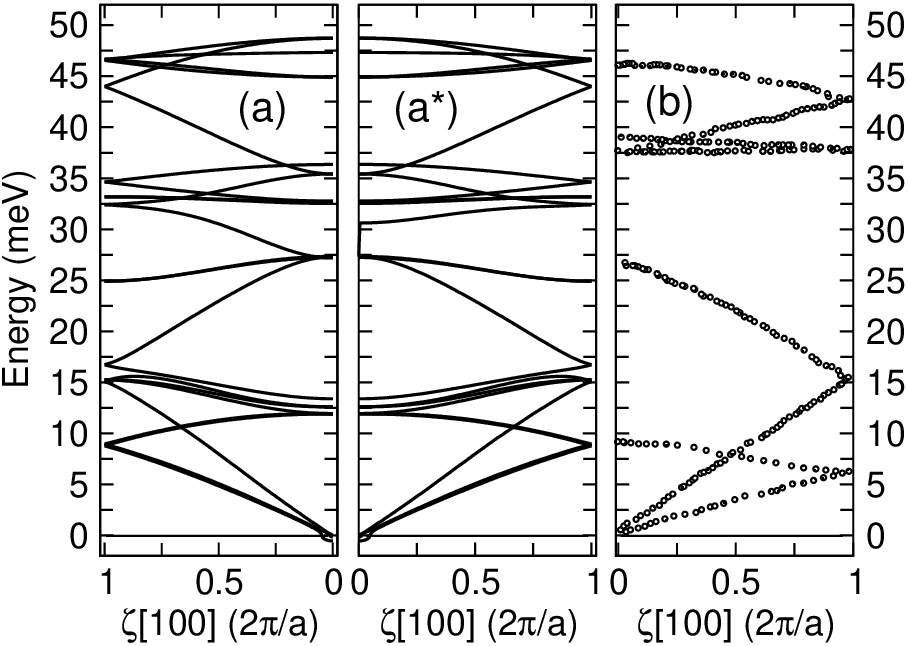}}
 \caption{Adeagbo et al.}
  \label{Tran}
\end{figure}

\begin{figure} 
 \centering
\resizebox{14cm}{!}{\includegraphics*{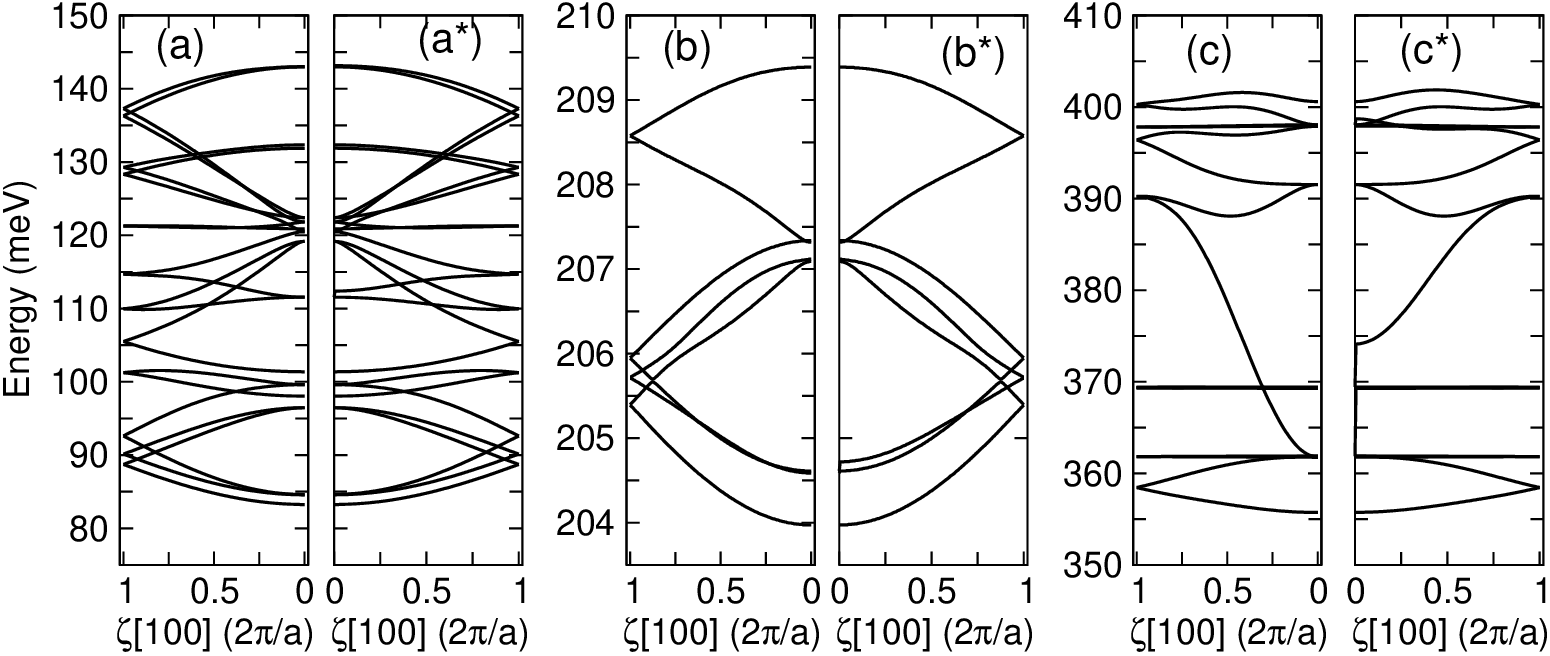}}
\caption{Adeagbo et al.}
  \label{libest}
\end{figure}

\begin{list}{}{\leftmargin 2cm \labelwidth 1.5cm \labelsep 0.5cm}

\item[\bf Fig.~\ref{cellice}] 
  Initial and the relaxed geometry of the
  unit cell of ice. The ice structure was initially packed in a cubic
  unit cell with initial lattice constant taken from the
  literature~\cite{Lee1993} to be 6.35 {\AA}. There are no
  hydrogen bonds in the initial prepared structure shown on the left
  but were perfectly formed after the relaxation according to 
  the GGA calculation. The relaxed geometry has the values of 
  $a \approx b \ne c$ which implies that the relaxed structure 
  is tetragonal with ${c}/{a}$ ratio $\approx$ 0.988

\item[\bf Fig.~\ref{dispVDOS}] 
Calculated dispersion curves for ice (a) with  no effective 
charges $Z^*$ taken into account and (b) with $Z^*$ taken into account.  
The curve on the left is the integrated phonon density of states. The 
frequencies $\nu_1$,  $\nu_2$, are  $\nu_3$  are respectively bending, 
symmetric and anti-symmetric  stretching analogous to the vibrational 
mode of an isolated water molecule~\cite{Adeagbo2005}.     
Note that the vibrational phonon density of states is not complete as it 
requires summation over all points in the first Brillouin zone, nevertheless 
the calculated G($\omega$) for both with and  without $Z^*$ do not differ.

\item[\bf Fig.~\ref{Tran}] 
Phonon dispersion in the transitional region 
showing the splitting of LO mode in the translational region. The dispersions 
are compared for the cases of both with and without the dipole interaction 
through the calculation of effective charges $Z^*$. The case with $Z^*$ 
is marked by a$^*$. The experimental dispersion~(\cite{Renker1973} taken from 
~\cite{ Li1996}) is shown on the right

\item[\bf Fig.~\ref{libest}] 
Phonon dispersion in the librational, bending and 
stretching region showing the splitting of LO modes. The case with $Z^*$ 
is marked by a$^*$ as in Fig.~\ref{Tran}

\end{list}

\clearpage

Table\ {\bf I}: Adeagbo et. al
{\sf
\begin{verbatim}
   Dielectric constant in cartesian axis

   (   1.881163770   -0.000048586     -0.000019997 )
   (  -0.000048586    1.881154759      0.000112163 )
   (  -0.000019996    0.000112161      1.883382267 )

   Effective charges E-U in cartesian axis
 
    Water molecule (1)    (hydrogen atom 1)
   (        0.66721        0.37702        0.40166 )
   (        0.40921        0.53464        0.38200 )
   (        0.40785        0.36304        0.62434 )
    Water molecule (1)    (hydrogen atom 2)  
   (        0.66611       -0.37720        0.40208 )
   (       -0.40909        0.53493       -0.38272 )
   (        0.40784       -0.36353        0.62525 )
    Water molecule (2)    (hydrogen atom 1)         
   (        0.66619        0.37729       -0.40199 )
   (        0.40909        0.53496       -0.38262 )
   (       -0.40785       -0.36360        0.62507 )
    Water molecule (2)    (hydrogen atom 2)
   (        0.66694       -0.37691       -0.40154 )
   (       -0.40907        0.53471        0.38197 )
   (       -0.40768        0.36296        0.62440 )
    Water molecule (3)    (hydrogen atom 1)
   (        0.53461        0.40937       -0.38251 )
   (        0.37708        0.66692       -0.40207 )
   (       -0.36332       -0.40814        0.62483 )
    Water molecule (3)    (hydrogen atom 2)
   (        0.53459       -0.40938        0.38266 )
   (       -0.37714        0.66694       -0.40229 )
   (        0.36338       -0.40822        0.62507 )
    Water molecule (4)    (hydrogen atom 1)
   (        0.53516       -0.40898       -0.38225 )
   (       -0.37710        0.66634        0.40159 )
   (       -0.36333        0.40759        0.62478 )
    Water molecule (4)    (hydrogen atom 2)
   (        0.53515        0.40892        0.38210 )
   (        0.37715        0.66626        0.40141 )
   (        0.36335        0.40747        0.62434 )
    Water molecule (5)    (hydrogen atom 1)
   (        0.66666       -0.37682       -0.40144 )
   (       -0.40902        0.53480        0.38199 )
   (       -0.40761        0.36297        0.62446 )
    Water molecule (5)    (hydrogen atom 2)
   (        0.66648        0.37748       -0.40192 )
   (        0.40929        0.53519       -0.38256 )
   (       -0.40784       -0.36361        0.62469 )
    Water molecule (6)    (hydrogen atom 1)
   (        0.66694        0.37695        0.40158 )
   (        0.40919        0.53476        0.38206 )
   (        0.40781        0.36307        0.62443 )
    Water molecule (6)    (hydrogen atom 2)
   (        0.66639       -0.37738        0.40199 )
   (       -0.40926        0.53512       -0.38263 )
   (        0.40780       -0.36352        0.62484 )
    Water molecule (7)    (hydrogen atom 1)
   (        0.53456        0.40948       -0.38257 )
   (        0.37720        0.66723       -0.40225 )
   (       -0.36340       -0.40832        0.62487 )
    Water molecule (7)    (hydrogen atom 2)
   (        0.53464       -0.40928        0.38261 )
   (       -0.37704        0.66666       -0.40213 )
   (        0.36332       -0.40807        0.62503 )
    Water molecule (8)    (hydrogen atom 1)
   (        0.53514        0.40901        0.38225 )
   (        0.37720        0.66637        0.40161 )
   (        0.36345        0.40765        0.62469 )
    Water molecule (8)    (hydrogen atom 2)
   (        0.53515       -0.40887       -0.38209 )
   (       -0.37703        0.66620        0.40138 )
   (       -0.36322        0.40739        0.62441 )
    Water molecule (1)    (oxygen atom 1) 
   (       -1.08499       -0.00011       -0.07462 )
   (        0.00002       -0.96152        0.00068 )
   (       -0.09759        0.00062       -1.07021 )
    Water molecule (2)    (oxygen atom 2) 
   (       -1.08486       -0.00006        0.07446 )
   (       -0.00023       -0.96140        0.00063 )
   (        0.09737        0.00068       -1.07026 )
    Water molecule (3)    (oxygen atom 3) 
   (       -0.96205        0.00002       -0.00017 )
   (       -0.00008       -1.08534        0.07521 )
   (       -0.00013        0.09795       -1.06997 )
    Water molecule (4)    (oxygen atom 4) 
   (       -0.96115        0.00000       -0.00001 )
   (       -0.00008       -1.08473       -0.07392 )
   (        0.00006       -0.09639       -1.07014 )
    Water molecule (5)    (oxygen atom 5) 
   (       -1.08488       -0.00031        0.07430 )
   (       -0.00036       -0.96143        0.00038 )
   (        0.09733        0.00058       -1.07024 )
    Water molecule (6)    (oxygen atom 6) 
   (       -1.08501        0.00014       -0.07445 )
   (        0.00014       -0.96156        0.00039 )
   (       -0.09755        0.00049       -1.07018 )
    Water molecule (7)    (oxygen atom 7) 
   (       -0.96206       -0.00013       -0.00004 )
   (       -0.00004       -1.08536        0.07520 )
   (        0.00010        0.09795       -1.06999 )
    Water molecule (8)    (oxygen atom 8) 
   (       -0.96113        0.00011       -0.00001 )
   (        0.00005       -1.08471       -0.07392 )
   (       -0.00004       -0.09639       -1.07012 )
\end{verbatim}
}
\clearpage

\begin{figure}[!] 
\centering
\resizebox{12cm}{!}{\includegraphics*{adeagbo-fig-2.eps}}
\caption{Adeagbo et al.}
  \label{dispVDOS}
\end{figure}

\end{document}